\documentclass[10pt,prl,aps,twocolumn,showpacs,
superscriptaddress]{revtex4}
\usepackage{graphicx}
\begin{document}

\title{Striped phase in a quantum XY-model with ring exchange}

\author{A. W. Sandvik}
\affiliation{Department of Physics, {\AA}bo Akademi University, 
Porthansgatan 3, FIN-20500 Turku, Finland}
\affiliation{Department of Physics, University of California,
Santa Barbara, California 93106}

\author{S. Daul}
\altaffiliation{Present address: Swiss Re, Mythenquai 50/60, 8022 Z\"urich,
Switzerland.}
\affiliation{Institute of Theoretical Physics, University of California,
Santa Barbara, California 93106}

\author{R. R. P. Singh}
\affiliation{Department of Physics, University of California,
Davis, California 95616}

\author{D. J. Scalapino}
\affiliation{Department of Physics, University of California,
Santa Barbara, California 93106}

\date{\today}

\begin{abstract}
We present quantum Monte Carlo results for a square-lattice $S=1/2$ 
XY-model with a standard nearest-neighbor coupling $J$ and a four-spin 
ring exchange term $K$. Increasing $K/J$, we find that the ground state 
spin-stiffness vanishes at a critical point at which a spin gap opens 
and a striped bond-plaquette order emerges. At still higher $K/J$, this 
phase becomes unstable and the system develops a staggered magnetization. 
We discuss the quantum phase transitions between these phases.
\end{abstract}


\maketitle

Ring exchange interactions have for a long time been known to be present 
in a variety of quantum many-body systems \cite{thouless} and have been 
investigated rather thoroughly in solid ${\rm ^3He}$ \cite{helium}. They are 
also important for electrons in the Wigner crystal phase \cite{voelker,bernu}. 
In strongly correlated electron systems, such as the high-$T_c$ cuprates and 
related antiferromagnets, ring exchange processes are typically much weaker 
than the pair exchange $J$ \cite{shastry} and are often neglected. Four-spin 
ring exchange has, however, been argued to be responsible for distinct 
features in the magnetic Raman \cite{roger} and optical absorption spectra 
\cite{lorenzana}. Neutron measurements of the magnon dispersion 
have also become sufficiently accurate to detect deviations from the standard
pair exchange Hamiltonian (the Heisenberg model) and such discrepancies have 
been attributed to ring exchange \cite{coldea,matsuda}. Recently, ring 
exchange has attracted interest as a potentially important interaction that 
could lead to novel quantum states of matter, in particular 2D electronic spin 
liquids with fractionalized excitations
\cite{misguich,senthil,sachdev,sedgewick,demler,balents}. 
Furthermore, for bosons on a square lattice ring exchange has been shown to 
give rise to a ``exciton Bose liquid'' phase \cite{paramekanti}.

Here we study the effects of ring exchange in one of the most basic quantum 
many-body Hamiltonians---the spin-$1/2$ XY-model on a 2D square lattice. 
We use a quantum Monte Carlo method (stochastic series expansion, hereafter 
SSE \cite{sse1,sse2,sse3}) to study the low-temperature behavior of this 
system including a four-spin ring term. Defining bond and plaquette 
exchange operators
\begin{eqnarray}
B_{ij} & = & S^+_iS^-_j + S^-_iS^+_j = 2(S^x_iS^x_j + S^y_iS^y_j), 
\label{bond} \\
P_{ijkl} & = & S^+_iS^-_jS^+_kS^-_l + S^-_iS^+_jS^-_kS^+_l, 
\end{eqnarray}
the Hamiltonian is
\begin{equation}
H = -J\sum\limits_{\langle ij\rangle} B_{ij}
    -K\sum\limits_{\langle ijkl\rangle} P_{ijkl},
\label{ham}
\end{equation}
where $\langle ij\rangle$ denotes a pair of nearest-neighbor sites and
$\langle ijkl\rangle$ are sites on the corners of a plaquette. For $K=0$ 
this is the standard quantum XY-model, or, equivalently, hard-core bosons 
at half-filling with no interactions apart from the single-occupancy 
constraint. This system undergoes a Kosterlitz-Thouless transition at 
$T/J \approx 0.68$ \cite{loh,harada} and has a $T=0$ ferromagnetic moment 
$M_x=\langle S^x_i\rangle \approx 0.44$ \cite{xyt0,xyaws}. The $K$-term 
corresponds to retaining only the purely $x$- and $y$-terms of the full 
cyclic exchange.

\begin{figure}
\includegraphics[width=5.5cm]{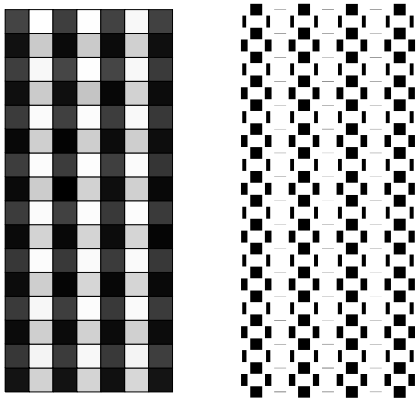}
\caption{Plaquette (left) and bond (right) strengths at the center of a
$64 \times 128$ open-boundary lattice at $K/J=10$ and $T=J/8$. 
The plaquette strengths 
are represented by shades of gray with the weakest $\langle P_{ijkl}\rangle 
=0.222$ (white squares) and strongest $0.468$ (black squares). The bond 
strengths are indicated by the width of the line segments, with the weakest
$\langle B_{ij}\rangle =0.181$ and the strongest $0.505$.}
\label{fig1}
\end{figure}

In a soft-core version of the pure ring model ($J=0$), Paramekanti {\it et al.}
recently found a compressible but non-superfluid phase (exciton Bose liquid) 
for weak on-site repulsion $U$ \cite{paramekanti}. As the hard-core limit is 
approached they found a transition to a staggered 
charge-density-wave phase. Hence, the ground state of the spin 
Hamiltonian (\ref{ham}) can be expected to change from an easy-plane 
ferromagnet with a finite spin stiffness $\rho_s$ and a magnetization 
$\langle M_x\rangle$ at low $K/J$ to an Ising-like antiferromagnet 
with vanishing $\rho_s$ and a staggered magnetization $\langle M_S\rangle 
=(-1)^{x_i+y_i}\langle S^z_i \rangle$ at large $K/J$. The central result of 
our simulations is that the competing  $J$ and $K$ 
interactions give rise to yet a third phase at $K/J \sim 10$; a striped 
bond-plaquette phase where the expectation values $\langle B_{ij}\rangle$ 
and $\langle P_{ijkl}\rangle$ alternate in strength with a period of 2 
lattice spacings in one of the lattice directions. An example of this order 
is illustrated in Fig.~\ref{fig1}. A similar columnar ``bond charge'' phase 
was recently predicted based on a lattice field-theory including a plaquette
term \cite{sachdev}. The field theory also has fractionalized phases, of 
which we have found no evidence. Hence, the microscopic mechanisms leading 
to fractionalized spin liquids remain to be clarified.

The SSE simulation method \cite{sse1,sse2,sse3} 
that we use here has previously been applied to a variety of spin and boson 
models with two-particle interactions, including the Hamiltonian (\ref{ham}) 
with $K=0$ (the XY-model) \cite{xyaws}. The generalization to include the 
four-spin $K$-term is relatively straight-forward, although non-trivial new 
procedures had to be developed for large-$K/J$ simulations \cite{method}.
Bond and plaquette strengths such as those shown in Fig.~\ref{fig1} 
were obtained using open-boundary rectangular $L_x\times L_y$ lattices with 
$L_y=2L_x$. The translational and rotational symmetries are then broken and 
a unique static bond-plaquette strength pattern can be observed when $K/J 
\sim 10$ at $T/J \alt 0.5$. For $K/J \alt 8$ no order is visible at the 
centers of large lattices at any temperature. The modulations seen within 
the stripes in Fig.~\ref{fig1} are strongest at the four corners of the 
lattice and decrease as the center is approached. They also decrease as the 
lattice size is increased and in the thermodynamic the striped state should 
therefore be analogous to the four-fold degenerate columnar spin-Peierls 
state of Ref.~\onlinecite{sachdev}.

Our conclusion that the stripes are stable is based on finite-size scaling
of correlation functions on periodic $L\times L$ lattices. The striped phase 
can be detected using the bond or plaquette correlations. Here we consider 
the plaquette structure factor
\begin{equation}
P(q_x,q_y) = {1\over L^2}\sum\limits_{a,b} 
{\rm e}^{i({\bf r}_a-{\bf r}_b)\cdot {\bf q}}
\langle P_{a_1a_2a_3a_4}P_{b_1b_2b_3b_4} \rangle ,
\end{equation}
where $a_1,\ldots,a_4$ are the sites belonging to plaquette $a$. We have 
studied the full ${\bf q}$-dependence and only found peaks at $(0,\pi)$ 
and $(\pi,0)$. Hence, the modulations within the stripes seen in 
Fig.~\ref{fig1} are indeed induced by open boundaries. The spin structure 
factor is defined as
\begin{equation}
S(q_x,q_y) = {1\over L^2}\sum\limits_{j,k} 
{\rm e}^{i({\bf r}_j-{\bf r}_k)\cdot {\bf q}}
\langle S^z_jS^z_k \rangle ,
\end{equation}
where $r_i=(x_i,y_i)$ is the lattice coordinate. 
We will analyze the staggered and striped order parameters per site,
defined as
\begin{eqnarray}
\langle M^2_S\rangle & = & S(\pi,\pi)/L^2 ,\\
\langle M^2_P\rangle & = & P(\pi,0)/L^2.
\end{eqnarray}
The spin stiffness (the superfluid density in the boson representation) is 
defined by
\begin{equation}
\rho_s = {\partial ^2 E(\phi) \over \partial\phi ^2},
\label{rho}
\end{equation}
where $E(\phi) = \langle H(\phi)\rangle /L^2$ and the twist $\phi$ is
imposed in the $x$ or $y$ direction so that the corresponding bond operators 
(\ref{bond}) become $B_{ij}(\phi)=\cos{(\phi)}(S^x_iS^x_j  + S^y_iS^y_j) +
\sin{(\phi)}(S^x_iS^y_j  - S^y_iS^x_j)$. 
The derivative at $\phi=0$ in Eq.~(\ref{rho}) can be directly estimated using
the winding number fluctuations in the SSE simulations \cite{sse4}.

\begin{figure}
\includegraphics[width=8.4cm]{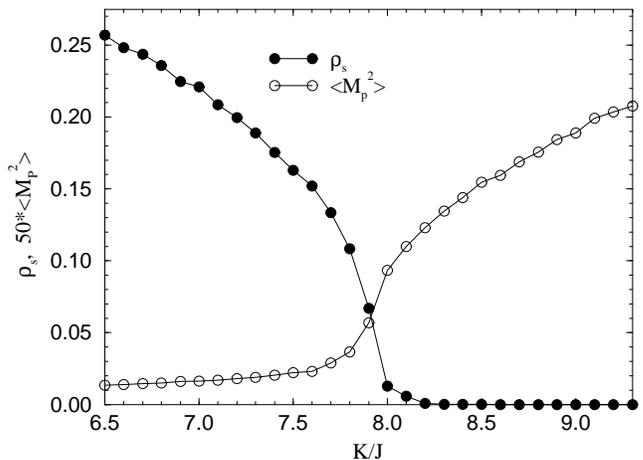}
\caption{Spin stiffness and plaquette-stripe order parameter vs
ring-exchange coupling for a $64\times 64$ system with periodic
boundary conditions at $T=J/8$.}
\label{fig2}
\end{figure}

\begin{figure}
\includegraphics[width=8cm]{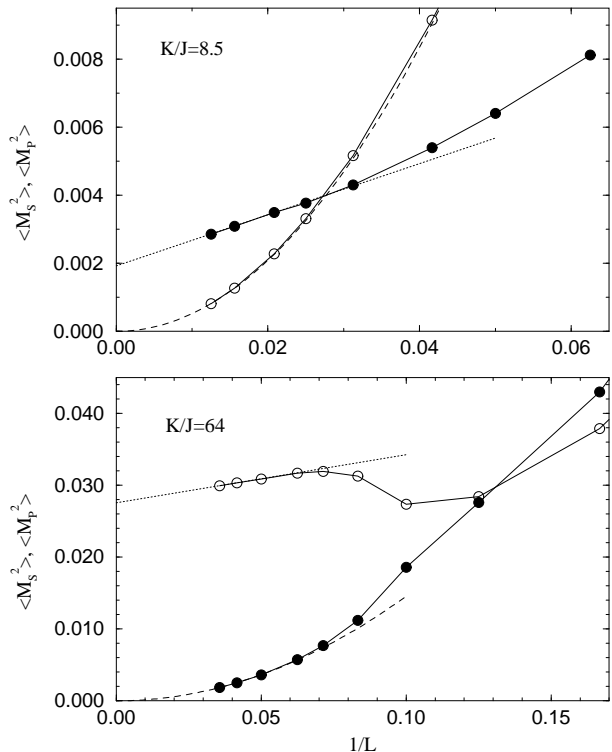}
\caption{Finite-size scaling of the ground state staggered magnetization 
(open circles) and the plaquette-stripe order parameter (solid circles) at 
$K/J=8.5$ and $64$. The dotted straight lines show extrapolations of the 
infinite-size order parameters. The dashed curves show the form $\sim 1/L^2$ 
expected asymptotically when there is no long-range order.}
\label{fig3}
\end{figure}

Fig.~\ref{fig2} shows the spin stiffness and the stripe order 
parameter on an $L=64$ lattice at $T=J/8$ (where the results are almost 
converged to their ground state values). The stiffness becomes very small at 
$K/J \approx 8$, where the stripe order increases significantly. Finite-size 
scaling shows that the stripe order 
survives in the thermodynamic limit. Results for $K/J=8.5$ and temperatures
sufficiently low to give the ground state are shown in Fig.~\ref{fig3}. 
For $L \agt 32$ the data graphed versus $1/L$ fall on a straight line, which 
extrapolates to a non-zero value as $L\to \infty$. Based on results 
\cite{paramekanti} for the soft-core version of the  $J=0$ model (or $K \to 
\infty$) the staggered magnetization can be expected to be non-zero for large
$K$. However, as also shown in Fig.~\ref{fig3}, at $K/J=8.5$ 
$\langle M_S^2\rangle$ decreases as $1/L^2$ for large lattices, implying that
the spin-spin correlations are short ranged [$S(\pi,\pi)$ is finite]. 
Fig.~\ref{fig3} also shows results for $K/J=64$, where the scaling behaviors 
of the two quantities is reversed---$\langle M_P^2\rangle$ decays as $1/L^2$ 
whereas $\langle M_S^2\rangle$ extrapolates to a non-zero value. Note that 
the size-dependence of $M_S^2$ is non-monotonic, 
with a minimum around $L \approx 10$. Such non-monotonicity has previously 
been observed for a spatially anisotropic spin model \cite{mchain} where it 
was attributed to the presence of two different low-energy scales in the 
system. The non-monotonicity seen at $K/J=64$ in Fig.~\ref{fig3} indicates 
that the stripe correlations remain strong with a correlation length $\sim 10$
lattice spacings. The location of the minimum in $\langle M_S^2\rangle$ moves 
to lower $1/L$ as $K/J$ is decreased, indicating growing stripe correlations. 
The strong stripe correlations in the staggered phase makes it difficult to 
determine the $\langle M_S \rangle$ versus $K/J$ curve. Our simulations 
show that the stripe order persists at least for $K/J$ up to $12$, and also 
that the staggered correlations are short ranged up to this coupling. The 
stripe correlations are short ranged for $K/J \ge 16$. Between $K/J=12$ 
and $16$ the two phases could either co-exist or be separated by a 
first-order transition. Simulations of larger lattices will be required 
in order to clarify the interesting transition region. 

\begin{figure}
\includegraphics[width=7.5cm]{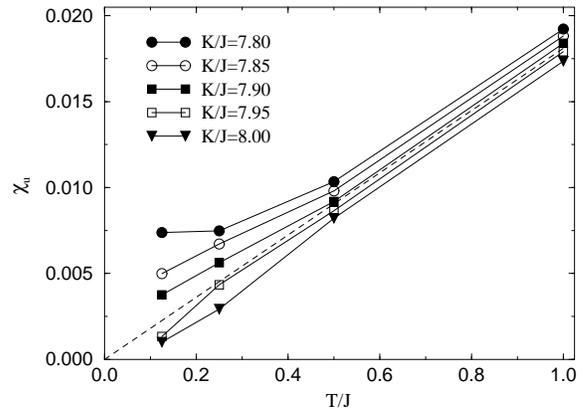}
\caption{Temperature dependence of the uniform magnetic susceptibility
for $L=80$ systems close to the superfluid-striped transition. Statistical 
errors are of the order of the size of the symbols. The dashed line shows the
linear behavior expected for a quantum phase transition with $z=1$.}
\label{fig4}
\end{figure}

The superfluid-striped transition appears to be of second order, although 
we cannot exclude a very weakly first-order transition (which was argued to 
be more likely in Ref.~\onlinecite{sachdev}). The vanishing of the 
spin stiffness seen in Fig.~\ref{fig2} indicates the opening of a spin gap.
A spin gap can be inferred also from the temperature dependence of the 
uniform magnetic susceptibility,
\begin{equation}
\chi_u = {1\over  L^2}{1\over T} 
\left \langle \left ( \sum\limits _i S^z_i \right )^2 \right \rangle .
\end{equation}
Fig.~\ref{fig4} shows the $T$-dependence for $L=80$ (sufficiently 
large to eliminate finite-size effects). The $T\to 0$ susceptibility 
vanishes for $K/J$ between $7.90$ and $7.95$, i.e., a spin gap is present 
above a critical coupling in 
this range. The temperature independence of $\chi_u$ at $K/J=7.80$ at the 
two lowest temperatures is expected on account of this being the behavior 
in the XY-model \cite{hasenfratz,xyaws,sse2}. The behavior for $K/J =  7.90$ 
and $7.95$ is consistent with $\chi_u \sim T$ at the critical coupling, 
which is indicative of a $T=0$ quantum critical point with dynamic exponent 
$z=1$ \cite{chn,chubukov}. At $K/J \approx 7.9$ we have verified that the 
stripe structure factor indeed exhibits non-trivial finite-size scaling, 
$P(\pi,0) \sim L^\epsilon$ with $\epsilon < 2$, but the statistical accuracy 
is not sufficient for determining the exponent to a meaningful precision. 
Nevertheless, power-law scaling for the same $K/J$ at which the spin gap 
opens supports a continuous quantum phase transition with no intervening 
disordered phase or co-existence region.

In summary, the spin-$1/2$ XY-model with ring exchange exhibits three 
different ground state orderings as a function of the strength of the ring 
term. The superfluid-striped transition appears to be a continuous quantum
phase transition, whereas the striped-staggered transition most likely is 
of first order. Since the sign of the $J$-term in (\ref{ham}) is irrelevant
(the sign of the $K$-term is relevant) the superfluid-striped transition
could possibly, in an extended parameter space, connect to the order-disorder
transition in the two-dimensional Heisenberg antiferromagnet with frustrating
interactions \cite{read}. We also note that the staggered-striped-superfluid
phase behavior versus $J/K$ shows interesting similarities to the high-$T_c$ 
cuprates, where the pseudogap phase intervening between the antiferromagnetic
and superconducting phases exhibits strong stripe correlations \cite{hightc}. 
Although the microscopic physics and symmetries are clearly different, a 
detailed study of the staggered-striped transition may still be useful 
in this context.

In spite of the absence of a spin liquid phase, the presence of three distinct
ordered ground states, and the phase transitions between them, puts the $J-K$ 
model (\ref{ham}) in an important class among the basic quantum many-body
Hamiltonians. Although the interesting large-$K/J$ region may not be of direct
relevance to real systems, we expect this and related model to be very useful 
as systems where complex quantum states and quantum phase transitions can be 
further explored on large lattices without approximations. Although other 
models, such as the frustrated $J_1-J_2$ Heisenberg model \cite{read}, may 
show similar or potentially even more complex behavior, sign problems 
affecting quantum Monte Carlo makes it difficult to obtain conclusive 
results. It would clearly be interested to study also the $J-K$ model with 
a positive sign for the $K$-term, in particular to determine whether
fractionalized spin liquid phases could arise, but unfortunately this also 
leads to sign problems.

The $J-K$ model with the sign of $K$ chosen here can be modified in several 
interesting ways and still be easily accessible to simulations using the 
SSE method. For example, when relaxing the hard-core constraint there should 
be a transition to a exciton Bose liquid phase \cite{paramekanti}, both as a 
function of on-site repulsion $U$ for large $K/J$ and as a function of $K/J$.
It will also be interesting to include a magnetic field to ``dope'' the 
striped and staggered phases. Transitions between different charge-density 
phases and the question of the existence of doped supersolid phases have 
recently been studied numerically for boson models where charge-density 
phases are stabilized due to diagonal density-density interactions 
\cite{hebert}. In contrast, the striped phase found here arises out of a 
competition between two kinetic terms and it may hence behave 
differently upon doping.

We would like to thank L. Balents, M.~P.~A. Fisher, S. Kivelson, and
A. Paramekanti for valuable discussions. We acknowledge support from
the Academy of Finland and the V\"ais\"al\"a Foundation (AWS), the 
Department of Energy, Grant No. DE-FG03-85ER45197 (DJS), and the National 
Science Foundation, Grant No. DMR-9986948 (RRPS). AWS would also like to 
thank L. Balents for Packard Foundation support for a visit to UCSB. 

\null


\begin{thebibliography}{00}
\bibitem{method}
In order to implement the SSE ``operator-loop'' algorithm \cite{sse2} 
when the $K$-term is included, all operators in the Hamiltonian are expressed 
as plaquette operators, i.e., a bond-operator $B_{ij}$ is written as 
$B_{ij}I_{kl}$, where $I_{kl}$ is a unit operator. Constant operators 
$cI_{ijkl}$ are introduced in order to carry out the SSE ``diagonal update''. 
The operator-loops are one-dimensional closed operator paths along which 
substitutions of the types $cI_{ijkl} \leftrightarrow B_{ij}I_{kl}$ and 
$B_{ij}I_{kl} \leftrightarrow P_{ijkl}$ are accomplished. The performance 
can be optimized using ``directed loop'' detailed balance equations of the 
type introduced in Ref.~\onlinecite{sse3}. The above scheme does not work 
when $J=0$ and is inefficient for large $K/J$. A new type of cluster update 
effecting substitutions of the type $P_{ijkl} \leftrightarrow cI_{ijkl}$ 
has been developed in order to study large $K/J$. Further details of the 
simulation algorithm will be presented elsewhere.
\end{thebibliography}
\end{document}